\documentclass[twocolumn,aps,prl,showpacs]{revtex4}

\begin{document}
\title{Reply to Marinatto's comment on ``Bell's
theorem without inequalities and without probabilities for two
observers''}
\author{Ad\'{a}n Cabello}
\email{adan@us.es} \affiliation{Departamento de F\'{\i}sica
Aplicada II, Universidad de Sevilla, 41012 Sevilla, Spain}
\date{\today}


\begin{abstract}
It is shown that Marinatto's claim [Phys. Rev. Lett. {\bf 90},
258901 (2003)] that the proof of ``Bell's theorem without
inequalities and without probabilities for two observers'' [A.
Cabello, Phys.~Rev.~Lett. {\bf 86}, 1911 (2001)] requires four
spacelike separated observers rather than two is unjustified.
\end{abstract}
\pacs{03.65.Ud,
03.65.Ta}

\maketitle


In his Comment~\cite{Marinatto03}, Marinatto claims that the proof
of Bell's theorem without inequalities in~\cite{Cabello01a}
requires four spacelike separated observers rather than two, as
asserted in~\cite{Cabello01a}. Marinatto's claim is based on the
fact that to test some of the properties used in the proof, for
instance the property
\begin{equation}
P_\psi (B_2=B_4 | A_1 A_3=+1) = 1,
\label{Alice1}
\end{equation}
one of the observers (Bob) must measure the spin along one
direction of his particle~$2$, $B_2$, and also the spin along one
direction of his particle~$4$, $B_4$. Marinatto argues that, since
both measurements are not spacelike separated, then
measuring~$B_2$ could disturb~$v(B_4)$ (the element of reality
corresponding to~$B_4$), and measuring~$B_4$ could disturb
$v(B_2)$ (the element of reality corresponding to~$B_2$).
Therefore, he maintains that, to avoid such possible disturbances,
both measurements should be spacelike separated.

However, such a prevention is not needed, because it can be
demonstrated that measuring~$B_2$ does not disturb~$v(B_4)$, and
measuring~$B_4$ does not disturb~$v(B_2)$. Let us recall Einstein,
Podolsky, and Rosen's (EPR) criterion for elements of reality:
{\em ``If, without in any way disturbing a system, we can predict
with certainty (i.e., with probability equal to unity) the value
of a physical quantity, then there exists an element of physical
reality corresponding to this physical quantity''}~\cite{EPR35}.
In the scenario described in~\cite{Cabello01a}, Alice, by means of
a spacelike separated measurement of the spin along one direction
of her particle~$1$, $A_1$, can use the property
\begin{equation}
P_\psi (A_1=B_2) = 0
\end{equation}
to predict with certainty the element of reality~$v(B_2)$.
Analogously, she, by means of a spacelike separated measurement of
$A_3$ on her particle~$3$, can use the property
\begin{equation}
P_\psi (A_3=B_4) = 0
\end{equation}
to predict with certainty the element of reality~$v(B_4)$. Note
that, according to EPR's criterion, what allows Alice to conclude
that there is an element of reality, for instance~$v(B_2)$, is the
fact that the result of Bob's measurement of~$B_2$ {\em can be
predicted with certainty}. The fact that close to particle~$2$
there could exist (or not) a second particle (in this case
particle~$4$) on which Bob could perform one measurement or other
does not invalidate Alice's prediction, since, according to the
predictions of quantum mechanics (presumably corroborated by any
conceivable experiment), the presence (or absence) of particle~$4$
close to particle~$2$, or the fact that Bob performs one
experiment or other on particle 4, do not change the result for
$B_2$ predicted by Alice (otherwise this effect could be used to
transmit information between spacelike separated regions).
Therefore, I must conclude that Marinatto's argument does not
justify the need for more spacelike separated observers. Indeed,
the proof of Bell's theorem without inequalities using four
qubits~\cite{Cabello01a} can be transformed into a proof using
only two particles~\cite{Cabello03,CPZBZ03}.



\end{document}